\documentstyle[12pt]{article}
\begin{document}
\baselineskip=20pt

\title{Remark on the variational principle in the 
AdS/CFT correspondence for the scalar field}

\author{\large Henrique Boschi-Filho\footnote{\noindent e-mail: 
boschi @ if.ufrj.br}\,  
and 
Nelson R. F. Braga\footnote{\noindent e-mail: braga @ if.ufrj.br}
\\ 
\\ 
\it Instituto de F\'\i sica, Universidade Federal  do Rio de Janeiro\\
\it Caixa Postal 68528, 21945-970  Rio de Janeiro, RJ, Brazil}
 
\date{}

\maketitle

\vskip 3cm

\begin{abstract}
We discuss how the  variational principle  can be used as a criterion
for choosing, among scalar field actions implying the same equation 
of motion, the appropriate one for the AdS/CFT  correspondence. 
\end{abstract}


\vfill\eject

\section{Introduction} 
The important physical consequences of Maldacena conjecture \cite{Malda} 
on the equivalence (or duality) of the large $N$ limit of $SU(N)$ 
superconformal field theories in $n$ dimensions and supergravity on 
anti de Sitter space-time in $n+1$ dimensions, $AdS_{n+1}$, attracted 
a lot of interest in the recent literature. By supergravity here, 
one  means the tree level approximation of  
string or M-theory defined on $AdS_{n+1} \times M_{d}$, 
where $M_{d}$ is some
$d$-dimensional compactification space.
This conjecture was further elaborated in the works of 
Gubser, Klebanov and Polyakov\cite{GKP} and Witten\cite{Wi}. 
In this, so called, AdS/CFT correspondence, field correlators of a 
conformal field 
theory on the $n$ dimensional boundary of an anti de Sitter space of 
dimension $\,n+1\,$ are defined by the dynamics of fields living inside 
this space. 
The boundary values of these fields, whose dynamics is defined in the bulk, 
have the role of sources for the field correlators of the conformal theory
on the boundary.\footnote{For recent reviews with a wide list of references 
see \cite{Pe} and \cite{Malda2}.} 

The presence of a boundary with non vanishing fields together with the 
fact that the metric is singular there may lead to some non standard 
results, comparing with usual quantum field theory, defined in spaces 
where the fields and field derivatives either vanish or have finite 
limiting values on the boundary.  
Classical actions that differ by total derivative terms (surface terms), 
having thus the same equations of motion, may play different roles in the 
AdS/CFT correspondence. 
In the case of fermionic fields, if one starts simply with the 
Dirac action as governing the dynamics inside $AdS_{n+1}$

$$ \int d^{n+1}x \sqrt{g} \,
{\overline \psi}\,({D\!\!\!\!/}-m)\, \psi\,
\,\,,\nonumber
$$

\noindent and imposes the equation of motion,
$({D\!\!\!\!/}-m)\, \psi\,=0$, one obtains a vanishing on shell
action. This would imply a trivial mapping between the bulk 
theory and the boundary correlators.
A solution to this problem was found by  Henningson 
and Sfetsos \cite{HS}, who showed that a non trivial mapping  
is possible if one introduces a surface term in the classical action.  
Soon after that, an interesting interpretation for the role played by this 
surface term was found by Henneaux \cite{He}, showing that the action 
is minimized by solutions of the equation of motion  
only if the surface term is added. 
Thus, the appropriate use of the variational 
principle furnishes, in the fermionic case a criterion for 
choosing the appropriate boundary term to be included 
in the action in the AdS/CFT correspondence.

One important point to be remarked is  the relation between the 
form of the fermionic action and the need for the surface term.
As the fermionic Lagrangian is linear in the field derivatives, the fields 
$\psi\,$ and $\,{\overline \psi}\,$ will be canonically conjugate 
variables.
So, as we are interpreting this fermionic fields as quantum operators one  
can not arbitrarily fix both of them on the boundary\cite{He}. That is why 
the action is minimized by the equation of motion only if one includes an 
additional surface term.
Things are different for the scalar field whose 
Lagrangian is quadratic in the field derivatives. In this case 
there is no problem in requiring that a solution of the equation of motion 
(just the field itself not the field derivatives) has some arbitrary 
limiting behavior in the boundary.
However, as we will see, the fact that one can not fix the field and
field time derivative simultaneously will make it possible again to use 
the variational principle as a guide for choosing the appropriate, 
among different actions that imply the same equation of motion 
in the bulk, for the AdS/CFT correspondence.
For massless scalar fields, if one chooses the standard form of the 
Klein Gordon action in AdS  (curved) spacetime
$${1\over 2} \int d^{n+1}x \sqrt{g} \,\partial_\mu \phi \,
\partial^\mu \phi
\,\,,\nonumber
$$

\noindent one does not need to introduce surface terms 
in the classical action in order to obtain the AdS/CFT correspondence, 
as we are going to review on section 2. 
If instead of this action, one chooses 
$$
 -\, {1\over 2} \int d^{n+1}x \sqrt{g} \phi \,\nabla_\mu 
\,\nabla^\mu \phi\,\,,\nonumber
$$

\noindent (where $\nabla_\mu$ is the covariant derivative in AdS space)  
one would get a vanishing generating functional for the correlators on 
the boundary. 
Both actions have the same Euler Lagrange equation of motion
$\,\nabla_\mu \,\nabla^\mu \phi \,=\,0$, and would thus describe  
the same dynamics in spaces 
with fields and field derivatives 
vanishing on the boundary.
However only the first one gives the appropriate mapping in the AdS/CFT 
correspondence. 
The aim of this letter is to show how the variational principle can 
be used also in the scalar field case so as to select  the appropriate 
action in the AdS/CFT context.  
In section {\bf 2} we review the AdS/CFT correspondence in the scalar 
field case and also make some remarks about the regularizations that 
must be introduced in order to have a well defined meaning to some singular 
quantities that show up, in particular in the discussion of the 
variational principle. In section {\bf 3} we discuss the case of 
the (would be) partially integrated action, show why this action is 
not minimized by the equations of motion and extend 
these results for the case of massive fields. Then,  
in section {\bf 4} we present some concluding remarks. 
We also included an appendix where we describe the behavior of the field 
time derivative which is useful in our discussion of the variational 
principle.


\section{Massless Scalar Fields on AdS/CFT 
and the Variational Principle}
 
Let us first consider a $(n+2)$-dimensional pseudo-Euclidean space-time 
with coordinates\footnote{We follow closely the 
notation and definitions of Petersen \cite{Pe}.}
 $(y^a)=(y^0,y^1,...,y^n,y^{n+1})$ and metric 
$\eta_{ab}={\rm diag}(+,-,-,...,-,+)$, so that the measure
\begin{equation}
y^2=(y^0)^2+(y^{n+1})^2-\sum_{i=1}^n(y^i)^2,
\end{equation}

\noindent is preserved under the transformations of the Lorentz group
SO(2,n). The $AdS_{n+1}$ can then be defined as a submanifold of this 
$(n+2)$-dimensional pseudo-Euclidean space-time such that
$y^2=b^2={\rm constant}\,.$ 
\noindent A pair of ``light cone'' coordinates can be defined as
\begin{eqnarray}
u=y^0+iy^{n+1}, \qquad v=y^0-iy^{n+1}
\end{eqnarray}

\noindent so that $y^2=uv-{\vec y}^2=b^2$. Then following Witten 
\cite{Wi}, we set $b = 0\,$ and define 
$( x^0,{\vec  x})\equiv(u^{-1},{\vec x})$, 
so that Anti de Sitter space-time can be characterized by the measure 
\begin{equation}
ds^2=\frac {(d x^0)^2}{( x^0)^2}+\frac{(d\vec x)^2}{( x^0)^2}\,
\equiv\frac 1{(x^0)^2}\sum_{\mu=0}^n(dx^\mu)^2
\end{equation}

\noindent 
in the upper half space $x^0\ge 0$ and $x^i\equiv x_i$, ($i=1,...,n$) 
are the coordinates in n-dimensional Euclidean space, $E^n$, 
in the boundary of AdS  space-time defined by $x^0\,=\,0\,$ plus 
a point at the infinity $x^0=\infty$.  
The metric in the $AdS_{n+1}\,$ space is taken as\footnote{\noindent  
Alternatively, the coordinates on the boundary could be 
defined to be an $n$-dimensional Minkowski space.}
\begin{equation}
\label{metric}
g_{\mu\nu}\,=\,{1\over (x^0)^2} \delta_{\mu\nu}\,\,,\,\,
\sqrt{g}\,=\,{1\over (x^0)^{n+1}}\,\,,\,\,
g^{\mu\nu}\,=\, (x^0)^2 \delta^{\mu\nu}\,\,.
\end{equation}

For massless scalar fields in the AdS/CFT correspondence one assumes 
that the field 
$\phi (x^0\,,\,x^i\,)$, where $i\,=\,1,...,n$ on $AdS_{n+1}\,$ 
has some boundary value 
$\phi_0 (\,x^i\,)$ as $x_0\,\rightarrow\,0\,$. 
This object will be coupled to a field 
${\cal O}\,(\,x^i\,)$, 
representing the CFT  on the  boundary. 
The correlation functions for the 
${\cal O}\,(\,x^i\,)$ field will then 
be calculated from the generator
\begin{equation}
Z\,[\,\phi\,]\,=\, \langle \,\,\exp \int d^{n}x\; \phi_0 (\,x^i\,)\,
{\cal O} (\,x^i\,)\,\,\rangle\,\,.
\end{equation}

\noindent where the integral is defined on the boundary $E^n$.

The mapping between the spaces is realized then by  associating this 
generator of CFT correlation functions on the boundary with the 
on shell value of the classical 
action of $AdS_{n+1}$.
One introduces the action that governs the dynamics inside the 
$AdS_{n+1}$ space as the 
standard Klein Gordon action in curved space-time: 
\begin{equation}
\label{action1}
I_1 [\phi ]\,=\, {1\over 2} \int d^{n+1}x \sqrt{g} 
\,\partial_\mu \phi \, \partial^\mu \phi
\,\,.
\end{equation}

\noindent The variation of the action $I_1$ when we change the field 
$\phi\,$ by a small amount $\,\delta\,\phi\,$ is formally
\begin{eqnarray}
\label{var1}
\delta I_1 [\phi ] &=& - \int d^{n+1}x\; \partial_\mu (\sqrt{g} \,
 \partial^\mu \phi\,) \,\delta \phi 
\,+\, \int d^{n+1} x\;  \partial_\mu \lbrack \sqrt{g} \,
( \partial^\mu \phi\,) \,\delta \phi \rbrack
\nonumber\\
&=&- \int d^{n+1}x\;  \sqrt{g} \,
( \nabla_\mu \partial^\mu \phi\,) \,\delta \phi 
\,+\, \int d^n x\;  \sqrt{g} \,
( \partial^0 \phi\,) \,\delta \phi 
\,\,.\label{vaction1}
\end{eqnarray}

\noindent However, the metric and also the field derivative, 
as we will see in the next section, are singular  at $x^0\,=\,0$.
Thus we need some prescription in order to have a well defined meaning to 
these expressions. We consider the last integral in (\ref{var1}) to be 
calculated near the boundary, at $x^0\,=\,\epsilon$. 
Moreover, the variations $\delta \phi\,$ of the fields in the bulk
are subject to the condition
\begin{equation}
\label{cond}
\left.\delta \phi\right|_{x^0 = \epsilon}\,=\,0\,\,.
\end{equation}

\noindent This condition does not mean that $\phi$ reaches the limiting
value $\phi_0$ at $x^0\,=\,\epsilon$ but rather that we are not 
varying the field configuration for $x^0\,< \,\epsilon$ with respect to the
classical solution.  Assuming these conditions to hold, the last integral
in (\ref{var1}) vanishes and thus the action is stationary with respect to 
variations of the field 
if the Euler Lagrange equation of motion is satisfied:
\begin{equation}
\label{motion}
\nabla_\mu \nabla^\mu \phi\,=\,{1\over \sqrt{g}} \partial_\mu 
\Big( \, \sqrt{g} \partial^\mu \phi \,\Big) 
\,=\,0\,\,.
\end{equation}

\noindent 
 
The field $\phi (x^0\,,\,{\vec x})\,$ inside the $AdS_{n+1}$ 
space is related  to the 
field on the boundary by calculating the Green's function that 
represents the evolution 
of the field from it's "initial" value at $x^0\,=\,0\,$ to 
the values inside 
$AdS_{n+1}$. Following \cite{Wi} one finds 
\begin{equation}
\label{solution}
\phi ( x^0\,,\,\vec{x}\,)\,=\,c\,\int d^n x^{\prime} 
{ (x^0)^n \over \left( (x^0)^2 \,+\,
(\vec x \,-\,{\vec x}^\prime \,)^2\,\right)^n} 
\phi_ 0 ({\vec x}^\prime )\,\,.
\end{equation}
The action $I_1[\phi]$, Eq. (\ref{action1}), with the metric 
(\ref{metric}) can be rewritten as
\begin{equation}
I [\,\phi\,]\,=\,{1\over 2} \int d^{n+1}x \, \partial_\mu 
\Big( ( x^0 )^{-n+1}\,
\phi \partial_\mu \phi \Big) \,-\, 
{1\over 2} \int d^{n+1}x \, \phi \, \partial_\mu 
\Big( ( x^0 )^{-n+1}\,
\partial_\mu \phi \Big) \,.
\end{equation}

So, considering the on shell value of the action, 
the last term vanishes by the 
equation of motion whereas the first one leads to a surface term. 
Using the Green function solution
(\ref{solution}) we can write the action as  
\begin{equation}
I [\phi ] \,=\, - {c n\over 2} \,\int d^n x \,d^n x^\prime 
{\phi_ 0 ({\vec x}^\prime )
\phi_ 0 ({\vec x} ) \over (\vec x \,-\,{\vec x}^\prime \,)^{2n}}
\end{equation}

\noindent that generates the appropriate two point functions for the 
operators ${\cal O} (\vec x )$  on the boundary:
\begin{equation}
\langle {\cal O} (\vec x ) {\cal O} ({\vec x}^\prime ) \rangle \sim
{ 1 \over (\vec x \,-\,{\vec x}^\prime \,)^{2n}  }.
\end{equation}


\section{Partially Integrated Scalar Field Action}

Now let us see how the previous discussion on AdS/CFT is modified 
if instead of action $I_1[\phi]$, Eq. (\ref{action1}), we decide 
to consider the action 
\begin{equation}
\label{action2}
I_2\, [\,\phi\, ]\,=\, -\, {1\over 2} \int d^{n+1}x \sqrt{g}\, 
\phi \,\nabla_\mu 
\nabla^\mu\, \phi\,\,.
\end{equation}

\noindent The Euler Lagrange equation of motion associated with 
this action is also given by $\nabla_\mu\nabla^\mu\phi=0$, 
Eq.~(\ref{motion}), {\sl i. e.}, it is the same as that 
corresponding to the action (\ref{action1}). 
In the standard field theory case, 
where one usually assumes that the space does not have a boundary, 
both actions would describe the same dynamics.

However, let us see what happens if we repeat the procedure described 
in the previous section in order to find the generator of correlation 
functions from action $I_2[\phi]$. 
Considering that the solution for the fields on $AdS_{n+1}$ 
in terms of the boundary values would
be the same Green function as given by eq. (\ref{solution}), 
since the equation of motion
for the fields described by $I_2[\phi]$ is the same as that for 
the previous action $I_1[\phi]$ action  we can rewrite
\begin{equation}
I_2\, [\,\phi\,]\,=\,-\, 
{1\over 2} \int d^{n+1}x \, \phi \, \partial_\mu 
\Big( ( x^0 )^{-n+1}\,
\partial_\mu\, \phi  \Big) \,,
\end{equation}

\noindent and this vanishes on shell. So we would find a vanishing 
generator of correlation functions if we use this action, 
analogously to the fermionic case discussed in the introduction. 

Let us now interpret this result in the light of the 
ideas of reference \cite{He} of 
appropriately using the variational principle to take 
into account the effects of 
the presence of the boundary.  
In the fermionic case the complete solutions for the fields 
$\psi \,$ and $\, {\overline \psi}\,$ can not have well defined 
values on the boundary of AdS .
The same thing does not happen in the scalar case, 
where we can assume the field to have  a well defined value 
$\,\phi_0\, $ on the boundary, at least for the massless case.  
There is however a difference in the action $I_2[\phi]$, 
Eq. (\ref{action2}), with respect to the previously considered action 
$I_1[\phi]$, Eq. (\ref{action1}), concerning the variational principle. 
The point is that the action $I_2$ involves second order field derivatives. 
For  field theories whose dynamics is governed by a classical action 
involving just first order field derivatives, the variational principle 
states that considering all field configurations with fixed values 
(of just the field itself) on a space time border, 
the action is minimized 
by the one that satisfies the Euler Lagrange equation. 
However, when one considers field theories involving higher order field 
derivatives one needs in general to fix also some of the derivatives of 
the fields on the border~\cite{CH,BB}.
Let us see what is the variation of action $I_2$ when  we make a small 
variation $\,\delta\phi\,$ in the field $\,\phi\,$:
\begin{eqnarray}
\label{vaction2}
\delta I_2 [\phi ] &=& 
- \int d^{n+1}x \lbrack \sqrt{g} \,
( \nabla_\mu \partial^\mu \phi\,) \,\delta \phi \rbrack
\,+\,  {1\over 2}
\int d^n x \lbrack ( x^0 )^{-n+1} \,
( \partial_0 \phi\,) \,\delta \phi \rbrack
\nonumber\\
& & - \int d^n x \lbrack (x^0)^{-n+1} \,
\phi   \delta ( \partial_0 \phi\,) \, \rbrack
\,\,.
\end{eqnarray}

So, that action would be stationary if, besides satisfying 
the equation of motion, we could 
make $\delta \phi$ and also $\delta \,(\,\partial_0 \phi\,)\,$ 
vanish on the boundary. 
This would correspond to look at solutions with well defined 
values of the field and also of its "time" 
derivative on the boundary. Assuming that the field has a well defined 
non vanishing boundary value, let us look at the derivative of $\phi$. 
Considering the field in the 
$AdS_{n+1}$ space as given by the Green function, Eq. (\ref{solution}),  
we calculate $\partial_0\,\phi$:
\begin{eqnarray}
\label{deriv}
\partial_0\; \phi ( x^0\,,\,\vec{x}\,) 
&=& {nc\over x^0} \,\int d^n x^{\prime} 
{ (x^0)^n \over \left( (x^0)^2 \,+\,
(\vec x \,-\,{\vec x}^\prime \,)^2\,\right)^n} 
\phi_ 0 ({\vec x}^\prime )
\nonumber\\
& & - \, 2nc \,\int d^n x^{\prime} 
{ (x^0)^{n+1} \over \left( (x^0)^2 \,+\,
(\vec x \,-\,{\vec x}^\prime \,)^2\,\right)^{n+1}} 
\phi_ 0 ({\vec x}^\prime )\,\,.\label{d0phi}
\end{eqnarray}

\noindent Choosing the normalization constant to be 
$c=(\sqrt{n}/\pi)^n$, 
we find (see the Appendix): 
\begin{eqnarray}
\label{sing}
\left.\partial_0 \; \phi ( x^0\,,\,\vec{x}\,)\right|_{x^0\to 0} 
= {n\over x^0} \, \phi_ 0 ({\vec x})
 - \, {2n\over x^0}\, \left({n\over n+1}\right)^{n\over 2} \,
 \phi_ 0 ({\vec x})\,\,,
\end{eqnarray}

\noindent so that the derivative of the field $\phi$ behaves as 
$(x^0)^{-1}\phi_0$ near the boundary.
 So, as we require the field $\phi$ to  have a well defined finite
value $\,\phi_0\,$ on the boundary, the derivative $\,\partial_0 \phi\,$
is not defined there essentially because of the singular nature of 
the metric at $x^0\to 0$. 
A regularized meaning to $\delta I_2[\phi]$, Eq. (\ref{vaction2}), 
is possible if consider again 
the surface integrals to be defined at $x^0\,=\,\epsilon\,$ and assume 
the condition $\delta\phi\vert_{x^0 =\epsilon}=0$, Eq. (\ref{cond}), 
to hold. If we were dealing with just classical objects there would 
be no objection to adding to (\ref{cond}) the extra  condition 
\begin{equation}
\label{cond2}
\left.\delta ( \partial_0 \phi ) \right|_{x^0 = \epsilon}\,=\,0\,\,.
\end{equation}

\noindent In this case the usual prescription of the variational 
principle including derivatives of the fields up to second order 
\cite{CH} would be satisfied and thus the action $I_2$ would be 
minimized by the solutions of the Euler Lagrange equations.
However, inserting the equation of motion in the action $I_2[\phi]$
we would find a vanishing boundary term contribution. So that this action
would lead to vanishing correlators in the  AdS/CFT correspondence.

The point is that we are considering the classical action just as an 
approximation for the
quantum action. Thus the field $\phi$ and its "time" derivative 
$\partial_0\phi$  are to be taken as quantum operators corresponding to 
a canonical pair of non commuting variables. 
So we can not impose that $\delta \phi$ and $\delta \partial_0\phi$ 
vanish at the same time, which would violate the uncertainty principle. 
As we are fixing the limiting value of the field, we can not assume 
the condition $\,\delta ( \partial_0 \phi\,)\,=\,0\,$ to hold 
on the boundary and the last term of equation (\ref{vaction2}) 
is non vanishing.
Therefore the action $I_2[\phi]$, Eq. (\ref{action2}), is not minimized 
by the configurations that satisfy the equation of motion
$\nabla_\mu\nabla^\mu\phi=0$, Eq. (\ref{motion}),
as long as we consider the field and its derivative as quantum operators.
If one decides, as in the fermionic case \cite{He}, to add to the action 
$I_2 $ a surface term that compensates for the variation of the action 
one  would find that the appropriate one is just
\begin{equation}
\label{surf}
M [\phi ]\,=\, \, {1\over 2} \int d^{n+1}x \sqrt{g} 
\,\nabla_\mu\,\Big(\,\phi \,\nabla^\mu \phi\,\Big)\,\,\,.
\end{equation}

That gives trivially $I_2\,+\,M\,=\,I_1$ showing then that  the  
analysis of the variational principle indicates that indeed the action 
$I_1$ is the appropriate one for calculating the field correlators on 
the boundary in the AdS/CFT correspondence.

It is interesting to mention that alternatively, one could have imposed 
that the derivative $\partial_0\phi$ to 
be well defined on the boundary with the price that the field $\phi_0$ 
itself would not be fixed, so one would still find non vanishing boundary 
terms. 
This alternative case should correspond to Neumann boundary conditions. 
It is even possible to discuss a generalized boundary condition where 
the condition is imposed on a linear combination of $\phi_0$ and 
$\partial_0\phi$. 
The calculations of the AdS  Green's functions for these two cases 
have been discussed recently by Minces and Rivelles \cite{Rivelles}.

Now, let us see if the presence of mass would change our 
conclusions. If we introduce a  mass term 
\begin{equation}
I_M [\phi]=\frac 12 \int d^{n+1}x\; m^2 \phi^2\,,
\end{equation}

\noindent into actions $I_1[\phi]$, Eq. (\ref{action1}), and 
$I_2[\phi]$, Eq. (\ref{action2}), the equation of motion 
is now is the massive Klein Gordon and one finds 
the solution \cite{Wi,Pe}
\begin{eqnarray}
\label{solution,m}
\phi ( x^0\,,\,\vec{x}\,)
&=& c\,\int d^n x^{\prime} 
{ (x^0)^{n+\lambda_+} \over \left( (x^0)^2 \,+\,
(\vec x \,-\,{\vec x}^\prime \,)^2\,\right)^{n+\lambda_+}} 
\phi_ 0 ({\vec x}^\prime )
\nonumber\\
&=& (x^0)^{-\lambda_+}\, c\,\int d^n x^{\prime} 
{ (x^0)^{n+2\lambda_+} \over \left( (x^0)^2 \,+\,
(\vec x \,-\,{\vec x}^\prime \,)^2\,\right)^{n+\lambda_+}} 
\phi_ 0 ({\vec x}^\prime )
\,\,,
\end{eqnarray}

\noindent where $\lambda_+$ is the highest root of 
$\lambda(\lambda+n)\,=\,m^2$. 

Note that the factor of $\phi_ 0 ({\vec x}^\prime )$
in the integrand of the above equation in the limit $x^0\to 0$
corresponds to an $n$-dimensional delta function, since
$$
\frac{\epsilon^\beta}{(\epsilon^2+{\vec x}^2)^\alpha}
$$
\noindent 
corresponds to $\delta^n(\vec x)$ up to a normalization, if one
guarantees that $2\alpha-n=\beta>0$. So we see that on the boundary,
$x^0\to 0$, the field $\phi (x^0,{\vec x})$ has the asymptotic behavior
\begin{equation}
\phi (x^0,{\vec x})\to (x^0)^{-\lambda_+}\,\phi_0 ({\vec x}).
\end{equation}

\noindent Now, the two point correlation functions for the 
operators ${\cal O}_\Delta (\vec x )$ with conformal dimension 
$\Delta=n+\lambda_+$  on the boundary are given by:
\begin{equation}
\langle {\cal O}_\Delta(\vec x ){\cal O}_\Delta({\vec x}^\prime )\rangle 
\sim { 1 \over (\vec x \,-\,{\vec x}^\prime \,)^{2n+2\lambda_+}  }.
\end{equation}

As the mass term does not involve field derivatives, there will be no 
extra surface term in $\delta I_2\,$. Then all the previous discussion 
about the variational principle remains the same and action $I_2$ is 
ruled out by the impossibility of fixing at the same time 
$\partial_0 \phi$ and $\phi$ (up to a scale factor) on the boundary. 

\section{Conclusions}

We have seen here that the idea of reference \cite{He} of using the 
minimum action principle as a criterion for selecting a classical action 
that will describe the mapping between the AdS  space and the conformal 
theory on the boundary also works properly in the scalar field  case.  
In contrast to the fermionic case, the field can be completely fixed
on the boundary. However, partially integrated actions involving the 
second order field derivative would be minimized only if one requires the 
additional condition of fixing the field derivative on the boundary. 
This condition is not possible for canonically conjugate quantum fields. 
So the variational principle, together with the quantum ingredient 
of the uncertainty principle rules out all the Lagrangians that 
would lead to  vanishing correlation functions.

\section*{Acknowledgments} 
The authors were partially supported by CNPq, FINEP and FUJB 
- Brazilian research agencies.

\section*{Appendix}

The integrand of the first term in (\ref{deriv}) behaves as an
 $n$-dimensional ``delta function'' times $\phi_0(\vec x^\prime)$ 
which implies a behavior like $(1/x^0)\phi(x^0,\vec x)$. 
The second term has the same behavior in the limit $x^0\to 0$, 
although this may be not transparent in the
above expression. To have a clue of this, let us recall a 
representation for the one-dimensional ``delta function''
\begin{equation}
\delta(x)={1\over\pi} \lim_{y\to 0} {y\over x^2+y^2}
\end{equation}

\noindent and consider (up to a constant) the product of rational 
functions in the $x^i\,$ variables
\begin{eqnarray}
&&{x^0\over (x^0)^2 + (x^1)^2 } \; 
{x^0\over (x^0)^2 + (x^2)^2 } \; \cdots \;
{x^0\over (x^0)^2 + (x^n)^2 }
\nonumber\\
&& \hskip 2cm =\;
{(x^0)^n\over (x^0)^{2n} 
+ (x^0)^{2(n-1)}\Big(\,(x^1)^2 + (x^2)^2 + \dots + (x^n)^2 \Big)
+ \; \dots}\,\,.
\end{eqnarray}

\noindent If we compare it with the 
term that shows up in (\ref{d0phi}) 
\begin{eqnarray}
&&{(x^0)^n\over\Big( (x^0)^2 + (x^1)^2 + (x^2)^2 + \dots
+ (x^n)^2 \Big)^n}
\nonumber\\
&&\hskip 2cm = \;
{(x^0)^n\over (x^0)^{2n} 
+ n(x^0)^{2(n-1)}\Big(\,(x^1)^2 + (x^2)^2 +\dots
+ (x^n)^2 \Big) 
+ \; \dots}\,,
\end{eqnarray}

\noindent where we are considering $x^i,\; (i=1,2,\dots,n)$ to be small
compared to $x^0$, since the ``delta functions'' are non vanishing
only for $x^1=x^2=\dots=x^n=0$, we find
\begin{eqnarray}
\lim_{x^0\to 0}\; {(x^0)^n\over\Big( (x^0)^2 + (x^1)^2 
+ (x^2)^2 + \dots
+ (x^n)^2 \Big)^n} \; = \;
(\pi)^n \delta(\sqrt{n}\,x^1) + \delta(\sqrt{n}\,x^2) 
\cdots \delta(\sqrt{n}\,x^n)\,,\nonumber\\
\end{eqnarray}

While, for the second term in eq. (\ref{d0phi}) we have
\begin{eqnarray}
&&{(x^0)^{n+1}\over\Big( (x^0)^2 + (x^1)^2 + (x^2)^2 + \dots
+ (x^n)^2 \Big)^{n+1}}
\nonumber\\
&&\hskip 1cm =\;
{(x^0)^{n+1}\over (x^0)^{2(n+1)} 
+ (n+1)(x^0)^{2n}\Big(\,(x^1)^2 + (x^2)^2 + \dots
+ (x^n)^2 \Big) + \dots}
\nonumber\\
&&\hskip 1cm =\;
{1\over x^0}\,{(x^0)^n\over (x^0)^{2n} 
+ (n+1)(x^0)^{2(n-1)}\Big(\,(x^1)^2 + (x^2)^2 + \dots
+ (x^n)^2 \Big) + \dots}\,\,.
\end{eqnarray}

\noindent Thus, taking the limit $x^0\to 0$ we obtain the product of
``delta functions'' 
\begin{eqnarray}
 &&\lim_{x^0\to 0} {(x^0)^{n+1}\over\Big( (x^0)^2 + (x^1)^2 
+ (x^2)^2 + \dots
+ (x^n)^2 \Big)^{n+1}}
 \nonumber\\ 
&&\qquad\qquad\qquad\qquad\qquad
\,\,=\,\,
{\pi^n\over x^0}\;
 \delta(\sqrt{n+1}\, x^1) 
 \delta(\sqrt{n+1}\, x^2) 
\cdots \delta(\sqrt{n+1}\, x^n)\;
 \nonumber\\ 
&&\qquad\qquad\qquad\qquad\qquad
\,\,=\,\,
{\pi^n\over x^0}\;
\delta^n(\sqrt{n+1}\, \vec x)\;,
\end{eqnarray}

\noindent where $\vec x=(x^1, x^2, \dots, x^n)$.
 In the general case 
$\epsilon^\beta/(\epsilon^2+{\vec x}^2)^\alpha$ 
the above formula can be extended to
\begin{eqnarray}
\lim_{\epsilon\to 0}
\frac{\epsilon^\beta}{{(\epsilon^2+{\vec x}^2)}^\alpha}
={\pi^n}\;
\delta^n(\sqrt{\alpha}\, \vec x)\;,
\end{eqnarray}

\noindent where $\vec x$ is a vector in $n$-dimensions 
as in the previous formula and  
the powers $\alpha$ and $\beta$ are such that 
$2\alpha -n=\beta>0$.



\end{document}